\documentclass[reprint, aip]{revtex4-1}
\usepackage[utf8]{inputenc}
\usepackage{graphicx, dcolumn, bm, textcomp, float, dblfloatfix}
\usepackage{physics, gensymb, natbib, hyperref, comment, array}

\begin{document}

\title{Using Quantum Annealers to Calculate Ground State Properties of Molecules}

\affiliation{Department of Physics and Astronomy, Purdue University}
\affiliation{Department of Chemistry, Purdue University}
\affiliation{Department of Mathematics, Purdue University}
\affiliation{Purdue Quantum Science and Engineering Institute, Purdue University}

\author{Justin Copenhaver}
\affiliation{Department of Physics and Astronomy, Purdue University}
\author{Adam Wasserman}
\affiliation{Department of Physics and Astronomy, Purdue University}
\affiliation{Department of Chemistry, Purdue University}
\affiliation{Purdue Quantum Science and Engineering Institute, Purdue University}
\author{Birgit Wehefritz-Kaufmann}
\affiliation{Department of Physics and Astronomy, Purdue University}
\affiliation{Department of Mathematics, Purdue University}
\affiliation{Purdue Quantum Science and Engineering Institute, Purdue University}

\begin{abstract}
Quantum annealers are an alternative approach to quantum computing which make use of the adiabatic theorem to efficiently find the ground state of a physically realizable Hamiltonian. Such devices are currently commercially available and have been successfully applied to several combinatorial and discrete optimization problems. However, the application of quantum annealers to problems in chemistry remains a relatively sparse area of research due to the difficulty in mapping molecular systems to the Ising model Hamiltonian. In this paper we review two different methods for finding the ground state of molecular Hamiltonians using Ising model-based quantum annealers. In addition, we compare the relative effectiveness of each method by calculating the binding energies, bond lengths, and bond angles of the H$_3^+$ and H$_2$O molecules and mapping their potential energy curves. We also assess the resource requirements of each method by determining the number of qubits and computation time required to simulate each molecule using various parameter values. While each of these methods is capable of accurately predicting the ground state properties of small molecules, we find that they are still outperformed by modern classical algorithms and that the scaling of the resource requirements remains a challenge.
\end{abstract}

\maketitle

\section{\label{Intro}Introduction}

The application of quantum computers to quantum chemistry has the potential to enable the simulation of large molecular systems that would otherwise be unattainable on classical computers. Thus far, several algorithms have been devised to simulate molecular systems on gate-based quantum computers, including the quantum phase estimation (QPE) and variational quantum eigensolver (VQE) algorithms \cite{Whitfield2010, Peruzzo2013}, as well as various quantum machine learning algorithms \cite{Xia2018, Choo2020}. While such algorithms show promise, the difficulty of constructing gate-based quantum computers has meant that their applications to quantum chemistry have been limited. Interest in quantum annealers as potential alternatives to gate-based quantum computers has increased recently, with new methods being proposed to map quantum chemistry problems onto quantum annealers \cite{Xia2017, Genin2019}. Here, we review some of the basics of quantum annealing and how it can be applied to the electronic structure problem, give a detailed explanation and comparison of recently developed mappings, and use these methods to calculate the ground state properties of the H$_3^+$ and H$_2$O molecules. In addition, we have open-sourced our code at \url{https://github.com/jcopenh/Quantum-Chemistry-with-Annealers} so that others can see how the techniques discussed might be implemented.

Quantum annealing is an optimization metaheuristic which exploits quantum tunneling effects to efficiently find the minimum of an objective function \cite{Kadowaki1998, Das2008, Falco2011}. The governing principle of quantum annealers is the adiabatic theorem, which states that if a system is in an eigenstate of a governing Hamiltonian which is slowly perturbed, then the system will remain in the instantaneous eigenstate of the perturbed Hamiltonian so long as the rate of change is slow enough and there is an energy gap between nearby eigenstates \cite{Ambainis2006}. The annealer begins in the ground state of some easy-to-prepare initial Hamiltonian $H_I$, and is then allowed to evolve approximately adiabatically whilst the governing Hamiltonian $H(t)$ is slowly perturbed to a final Hamiltonian $H_F$ according to

\begin{equation}
    H(t) = A(t) H_I + B(t) H_F
    \label{Annealing}
\end{equation}

\noindent
where the functions $A(t)$ and $B(t)$ are collectively referred to as the annealing schedule and satisfy the constraints $A(0) \gg B(0) \approx 0$ and $B(T) \gg A(T) \approx 0$, where $T$ is the total annealing time. \cite{Kadowaki1998, Amin2015}. At the end of the annealing process, $H(T)=H_F$ and the current state of the annealer is taken to be the ground state of $H_F$. Thus, by encoding information about a problem into the previously-unknown ground state of $H_F$, the annealer "solves" the problem by taking advantage of the adiabatic theorem to search for the ground state.

The prototypical example of a governing Hamiltonian is the transverse-field Ising model \cite{Kadowaki1998}:

\begin{equation}
    H(t) = A(t) \sum_i \sigma_x^i + B(t) \left[\sum_i h_i\sigma_z^i + \sum_{i<j} J_{ij}\sigma_z^i\sigma_z^j\right]
    \label{TransverseIsing}
\end{equation}

\noindent
where $h_i$ are the qubit biases, $J_{ij}$ the coupling coefficients, and $\sigma_{\alpha}^i$ are Pauli operators acting on the $i$th qubit of the annealer. The transverse-field term is used as the initial Hamiltonian $H_I$, leaving the remaining terms, which form a regular Ising spin-glass model, as the final Hamiltonian

\begin{equation}
    H_F = \sum_i h_i\sigma_z^i + \sum_{i<j} J_{ij}\sigma_z^i\sigma_z^j
    \label{Ising}
\end{equation}

The transverse-field Ising model is stoquastic, meaning all off-diagonal terms are real and non-positive. This is of great consequence for the capabilities of Ising model-based annealers, as any adiabatic quantum computer must implement a non-stoquastic Hamiltonian to be universal \cite{Biamonte2008, Vinci2017, Grant2020}. Indeed, previous experiments have shown annealers based on stoquastic Hamiltonians to be of limited applicability \cite{Cho2014, Amin2015}, leading to the development of several non-stoquastic annealer designs \cite{Barends2016, Vinci2017, Ozfidan2020}. However, given that such implementations are very much still in development, we will focus on the application of Ising model-based annealers in this paper.

The main challenge with utilizing Ising model-based quantum annealers to solve quantum chemistry problems lies in the difficulty of finding $H_F$ as given in Eq. \ref{Ising}. In the next section we give an overview of how the electronic structure problem can be mapped to an Ising model Hamiltonian.

\section{\label{MoltoIsing}Mapping Molecular Hamiltonians to the Ising Model}

The electronic structure of a molecule describes the motions and spin properties of electrons within the molecule. Descriptions of the electronic structure are typically given as solutions to the Schrödinger equation after applying the Born-Oppenheimer approximation, which assumes the nuclei to be fixed in space relative to the center of the molecule. The first-quantized Hamiltonian in atomic units is given by

\begin{equation}
\begin{split}
    H = -\sum_i\frac{\nabla_i^2}{2} - \sum_A\frac{\nabla_A^2}{2M_A}
        - \sum_{i,A}\frac{Z_A}{\lvert \boldsymbol{r}_i-\boldsymbol{R}_A \rvert} \\
        + \sum_{i<j}\frac{1}{\lvert \boldsymbol{r}_i-\boldsymbol{r}_j \rvert} + \sum_{A<B}\frac{Z_AZ_B}{\lvert \boldsymbol{R}_A-\boldsymbol{R}_B \rvert}
    \label{Electronic}
\end{split}
\end{equation}

\noindent
where $\boldsymbol{r}_i$ is the position of electron $i$ and $\boldsymbol{R}_A$, $M_A$, and $Z_A$ are the position, mass, and charge of nuclei $A$.

Using the second quantization formalism, one can write $H$ in terms of fermionic creation and annihilation operators $a_i^{\dagger}$ and $a_i$ by choosing a basis set, calculating the one-body and two-body integrals $h_{ij}$ and $h_{ijkl}$, and constructing $H$ as

\begin{equation}
    H = \sum_{i,j}h_{ij}a_i^{\dagger}a_j
        + \frac{1}{2}\sum_{i,j,k,l}h_{ijkl}a_i^{\dagger}a_j^{\dagger}a_ka_l 
    \label{Fermionic}
\end{equation}

\noindent
Eigenstates of the Hamiltonian are now given by their occupation-number representation,

\begin{equation}
    \ket{\Psi} = \ket{n_1, n_2,..., n_M}
    \label{OccupationNumber}
\end{equation}

\noindent
with $M$ being the total number of spin-orbitals in the chosen basis set and $n_i\in\{0,1\}$ representing whether or not spin-orbital $i$ is filled by an electron. The state $\ket{\Psi}$ is equivalent to the Slater determinant formed using the filled spin-orbitals. In addition, one can restrict the active space of the molecule to a subset of the spin-orbitals, reducing the number of fermionic operators in $H$ at the cost of potentially missing lower energy solutions. In this case, $M$ is now the number of spin-orbitals in the chosen active space.

Note that the second quantization formalism does not conserve particle number, i.e. the eigenstates of $H$ in Eq. \ref{Fermionic} are in the form of Eq. \ref{OccupationNumber} with anywhere from 0 to $M$ electrons filling the spin-orbitals. In many cases, however, one is only interested in solutions with a fixed number of electrons $N$. In order to ensure that the ground state solution for $H$ has $N$ electrons, one can construct the total number operator

\begin{equation}
    \hat{N} = \sum_{i=1}^Ma_i^{\dagger}a_i
    \label{NumberOperator}
\end{equation}

\noindent
and use it to add a penalty term to $H$:

\begin{equation}
    H' = H + w(N-\hat{N})^2
    \label{PenaltyTerm}
\end{equation}

\noindent
where $w$ is a weight factor large enough to ensure that the eigenvalues corresponding to solutions with $N$ electrons are less than all other eigenvalues \cite{Ryabinkin2019}. One can similarly add penalty terms to $H$ to fix the total spin of the system, or any other quantum observable, so long as one can readily construct the corresponding operator in terms of the creation and annihilation operators.

After writing $H$ in terms of creation and annihilation operators, one must transform $H$ into a sum of Pauli words:

\begin{equation}
    H = \sum_i\alpha_iP_i
    \label{PauliHamiltonian}
\end{equation}

\noindent
with the Pauli word $P_i$ being of the form

\begin{equation}
    P_i = \pm\{I, \sigma_x, \sigma_y, \sigma_z\}^{\otimes m}
    \label{PauliWord}
\end{equation}

\noindent
where $m \leq M$ is the number of qubits. Here, $H$ acts on the $m$-qubit space spanned by basis states of the form

\begin{equation}
    \ket{\phi} = \prod_{i=1}^m\ket{z_i} = \ket{z_1, z_2, ..., z_m}
    \label{QubitState}
\end{equation}

\noindent
such that $z_i\in\{0,1\}$ is equal to $0$ if qubit $i$ is in the spin-up state and $1$ if it is in the spin-down state. Any state $\ket{\psi}$ in this $m$-qubit space can then be written as a sum of these $2^m$ basis states:

\begin{equation}
    \ket{\psi} = \sum_{i=1}^{2^m}a_i\ket{\phi_i}
    \label{BasisStates}
\end{equation}

Several transformations exist to transform $H$ into the form of Eq. \ref{PauliHamiltonian}, including the Jordan-Wigner (JW) transformation, the Bravyi-Kitaev (BK) transformation, and the parity encoding, to name just a few \cite{BravyiKitaev2000, Seeley2012, Tranter2015}. Once this is done, it is often helpful to reduce the number of qubits required to simulate the molecule by exploiting its symmetries and conservation properties. A detailed explanation of this procedure can be found in Ref. \cite{BravyiGambetta2017}, along with a look into how such reductions can be found using knowledge of the molecule's point group symmetries in Ref. \cite{Setia2020}.

The next step in transforming $H$ into the form of Eq. \ref{Ising} is perhaps the most difficult. We are aware of two methods for mapping Eq. \ref{PauliHamiltonian} onto the Ising model Hamiltonian: the Xia-Bian-Kais (XBK) transformation proposed in Ref. \cite{Xia2017} and the Bloch angle mapping used in Ref. \cite{Genin2019}. These methods will be described in detail in sections \ref{XBK-Section} and \ref{QCC-Section}. Both will result in a diagonal Hamiltonian in the form of a k-local sum of z-type Pauli operators:

\begin{equation}
    H = \sum_i\alpha_i\sigma_z^i + \sum_{i<j}\alpha_{ij}\sigma_z^i\sigma_z^j
        + \sum_{i<j<k}\alpha_{ijk}\sigma_z^i\sigma_z^j\sigma_z^k + ...
    \label{k-local}
\end{equation}

\noindent
which now acts on the mapped $m'$-qubit space where $m' \geq m$.

The k-local Hamiltonian of Eq. \ref{k-local} must then be reduced to a 2-local Hamiltonian with the same ground state. This process, known as quadratization, in general leads to the introduction of several auxiliary qubits which account for the missing higher order terms \cite{Anthony2017, Dattani2019}. After quadratization, $H$ should now be in the form of Eq. \ref{Ising}, and can be embedded on the quantum annealer to find the ground state.

\section{\label{XBK-Section}Xia-Bian-Kais Method}

The XBK transformation maps states from the $m$-qubit space associated with Eq. \ref{PauliHamiltonian} to an $rm$-qubit space, where $r$ is a variational parameter which represents the number of copies of the original $m$ qubits \cite{Xia2017}. Each Pauli operator in this new space can be represented using tensor products of the identity and z-type Pauli operators. By increasing $r$, one expands the space in which the quantum annealer searches for the ground state, thus increasing the accuracy of the energy calculations.

The mapping of each Pauli operator to the new space is given by

\begin{equation}
\!
\begin{aligned}
\sigma_x^i &\rightarrow \frac{1 - \sigma_z^{i_j}\sigma_z^{i_k}}{2} & \sigma_y^i &\rightarrow \boldsymbol{i}\frac{\sigma_z^{i_k} - \sigma_z^{i_j}}{2} \\
\sigma_z^i &\rightarrow \frac{\sigma_z^{i_j} + \sigma_z^{i_k}}{2}  & I^i &\rightarrow \frac{1 + \sigma_z^{i_j}\sigma_z^{i_k}}{2}
\end{aligned}
\label{XBK}
\end{equation}

\noindent
with $\sigma_z^{i_j}$ being the z-type Pauli operator acting on $i$th qubit of the $j$th $m$-qubit subspace. For a given $i$ and $j$, applying Eq. \ref{XBK} to each operator in Eq. \ref{PauliHamiltonian} will map $H$ to a ``sub-Hamiltonian'' $H^{(i,j)}$ acting on a $2m$-qubit space. In order to properly account for each of the $\lceil\frac{r}{2}\rceil$ possible sign combinations of the sub-Hamiltonians, one defines a sign parameter $0 \leq p \leq \lfloor\frac{r}{2}\rfloor$ and constructs the sign function

\begin{equation}
    S_p(i) = \begin{cases}
        -1, & i \leq p \\
        1, & \text{else} 
    \end{cases}
    \label{rm-Hamiltonian}
\end{equation}

\noindent
The $\lceil\frac{r}{2}\rceil$ possible $rm$-qubit Hamiltonians are obtained by summing over $H^{(i,j)}$ for each combination of $1 \leq i,j \leq r$ and taking into account the signs associated with each sub-Hamiltonian:

\begin{equation}
    H'_p = \sum_{i,j \leq r}H^{(i,j)}S_p(i)S_p(j)
    \label{rm-Hamiltonian}
\end{equation}

\noindent
Each of the $H'_p$ will explore a sector of the $rm$-qubit space.

It can be shown that if the eigenvalue of the original $H$ corresponding to the $m$-qubit state $\ket{\psi}$ is $\lambda '$, then the eigenvalue of $H'_p$ corresponding to the $rm$-qubit state $\ket{\psi '}$ is $\lambda '\sum_ib_i^2$, where $b_i$ is the number of times the basis state $\ket{\phi_i}$ appears in $\ket{\psi '}$ \cite{Xia2017}. Thus, one can construct an operator $C_p$ which keeps track of $\sum_ib_i^2$,

\begin{equation}
    C_p = \sum_{\pm}\left[\sum_{i=1}^r\left(S_p(i)\prod_{k=1_i}^{m_i}\frac{1 \pm \sigma_z^k}{2}\right)\right]^2
    \label{C operator}
\end{equation}

\noindent
where $\sum_{\pm}$ means to sum over all combinations of the plus-minus signs. Given $H'_p$ and $C_p$, the procedure to find the minimum eigenvalue of $H$ for the $p$th sector is as follows: we choose a starting value $\lambda$ and construct the operator $D_{p,\lambda}=H'_p - \lambda C_p$, whose minimum eigenvalue is $\sum_ib_i^2(\lambda '-\lambda)$ so long as it is less than $0$. After quadratizating this operator we can map it to the quantum annealer by taking $H_F=D_{p,\lambda}$ to find $\sum_ib_i^2(\lambda '-\lambda)$. Evaluating $C_p$ at the ground state we obtain $\sum_ib_i^2$, which allows us to solve for $\lambda '$. Setting $\lambda = \lambda '$, we repeat this process until the minimum eigenvalue of $D_{p,\lambda}$ is greater than or equal to $0$. The minimum eigenvalue is then taken as $\lambda '$ when this process terminates.

By searching through all values of $p$, we find the minimum eigenvalue of $H$ for those states mapped to the $rm$-qubit space. To retrieve the $m$-qubit state $\ket{\psi}$ from the $rm$-qubit state $\ket{\psi '}$, we use the fact that for large enough $r$, the coefficient $a_i$ for the basis state $\ket{\phi_i}$ can be approximated by

\begin{equation}
    a_i \approx \frac{b_iS(b_i)}{\sqrt{\sum_jb_j^2}}
    \label{C operator}
\end{equation}

\noindent
where by $S(b_i)$ we mean the sign of the sum of $S_p(i)$ for all $m$-qubit spaces that are in the $i$th basis state, using the value of $p$ corresponding to the sector in which the ground state was found. We then use Eq. \ref{BasisStates} to obtain $\ket{\psi}$.

Beyond the pre-processing required to construct and quadratize $D_{p,\lambda}$, the optimization in the XBK method is performed solely on the quantum annealer. However, this pre-processing becomes quite expensive for larger values of $m$ and $r$. The number of qubits in $D_{p,\lambda}$ before quadratization is $rm$, and due to the large number of auxiliary qubits introduced during the quadratization procedure, the final number of qubits required to simulate $D_{p,\lambda}$ on the quantum annealer can quickly surpass what is available on modern systems. Thus, the application of the XBK method to the accurate simulation of larger molecular systems is currently impractical.

\section{\label{QCC-Section}Qubit Coupled Cluster Method}

The qubit coupled cluster (QCC) method is a hybrid classical-quantum method which utilizes the quantum annealer to improve the convergence rate of a classical optimization routine \cite{Genin2019, Ryabinkin2018}. The QCC method begins with the qubit mean-field (QMF) description, which assumes that the ground state of $H$ is of the form

\begin{equation}
    \ket{\psi} = \prod_{i=1}^m\ket{\Omega_i}
    \label{Ansatz}
\end{equation}

\noindent
such that $\ket{\Omega_i}$ is the spin-coherent state of the $i$th qubit:

\begin{equation}
    \ket{\Omega_i} = \cos(\frac{\theta_i}{2})\ket{0} + e^{\boldsymbol{i}\varphi_i}\sin(\frac{\theta_i}{2})\ket{1}
    \label{Coherent}
\end{equation}

\noindent
where $\varphi_i\in [0,2\pi)$ and $\theta_i\in [0,\pi)$ are the azimuthal and polar angles of the Bloch sphere. The set of all $\varphi_i$ and $\theta_i$ are collectively called the Bloch angles of $\ket{\psi}$. The QMF energy is then defined as the expectation value of $H$ evaluated at $\ket{\psi}$ for optimized Bloch angles.

The Hamiltonian in Eq. \ref{PauliHamiltonian} can be converted into a real-valued function whose global minimum is equal to the QMF energy by mapping the Pauli operators to the Bloch angles,

\begin{equation}
\!
\begin{aligned}
\sigma_x^i &\rightarrow \cos\varphi_i\sin\theta_i \\
\sigma_y^i &\rightarrow \sin\varphi_i\sin\theta_i \\
\sigma_z^i &\rightarrow \cos\theta_i
\end{aligned}
\label{BlochAngles}
\end{equation}

\noindent
With $H$ now in the form of a continuous optimization problem, a classical optimization routine can be used to find the QMF energy. Using the optimal Bloch angles, one can then use Eq. \ref{Coherent} to reconstruct the state corresponding to the QMF energy.

The next step of the QCC method is to introduce a multi-qubit unitary transformation to ``entangle'' the qubits and simulate electron-correlation effects. The QCC transformation is given by

\begin{equation}
    U(\boldsymbol{\tau}) = \prod_{k=1}^{N_{ent}}\exp(-\boldsymbol{i}\tau_kP_k/2)
    \label{QCC Transformation}
\end{equation}

\noindent
where $P_k$ is a multi-qubit Pauli word called an entangler, $\tau_k\in [0,2\pi)$ is the corresponding entangler amplitude, and $N_{ent}$ is the total number of entanglers used. As $N_{ent}$ is increased more electron-correlation effects are taken into account, improving the accuracy of the method. In addition, some entanglers will be more important in the simulation than others, necessitating a procedure to find the optimal entanglers for the system at hand as in Ref. \cite{Ryabinkin2018}. The transformed Hamiltonian $H'$ can easily be found using the recursive formula

\begin{equation}
\begin{split}
    H^{(k)}(\boldsymbol{\tau}) = H^{(k-1)} - \boldsymbol{i}\frac{\sin\tau_k}{2}[H^{(k-1)}, P_k] \\
    + \frac{1}{2}(1-\cos\tau_k)P_k[H^{(k-1)}, P_k]
    \label{Recursive}
\end{split}
\end{equation}

\noindent
for $1 \leq k \leq N_{ent}$, where $H^{(0)}=H$ and $H^{(N_{ent})}=U^{\dagger}HU=H'$. The QCC energy is defined as the expectation value of $H'$ for optimized Bloch angles and entangler amplitudes. Using the Bloch angle mapping given by Eq. \ref{BlochAngles}, $H'$ can be converted into a continuous optimization problem over the set of Bloch angles and entangler amplitudes, where the global minimum is now the QCC energy.

The quantum annealer is brought into the QCC method by recognizing the symmetries of the trigonometric functions present in $H$. The even-odd nature of these functions allows for their domains to be ``folded'' along their axis of symmetry by introducing discrete variables $Z_i\in\{-1,1\}$. These foldings turn $H$ into a mixed discrete-continuous optimization problem, which is solved in a step-based fashion. For fixed values of the continuous variables, $H$ will be in the form of Eq. \ref{k-local} and, after quadratization, can be mapped to the annealer as $H_F=H$. After using the annealer to optimize the discrete part, the classical computer is used to perform the continuous optimization with the discrete variables fixed.

By introducing the foldings, the chances of finding the QMF and QCC energies can be greatly improved at the cost of the discrete optimization step performed by the annealer. The foldings, which can be found in more detail in Ref. \cite{Genin2019}, allow for one folding in the $\theta_i$ variables, two foldings in the $\varphi_i$ variables, and two foldings in the $\tau_k$ variables. Thus, there are up to $3m$ discrete variables to be optimized when finding the QMF energy, and up to $3m+2N_{ent}$ discrete variables when finding the QCC energy. The number of qubits before quadratization is then equal to the number of discrete variables being optimized.

Unlike the the XBK method, the QCC method relies on a classical computer to perform the bulk of the optimization; the quantum annealer simply increases the chances of finding the correct minimal energy. Due to this reliance, the potential for a substantial improvement over other classical algorithms is dubious. However, with the correct choice of entanglers and foldings, the QCC method can produce results comparable to the XBK method whilst using fewer qubits on the annealer.

\section{\label{Results}Results}

In order to compare the relative accuracy of the XBK and QCC methods, we have used each method to calculate the binding energy and bond length of H$_3^+$ and the binding energy, bond length, and bond angle of H$_2$O. The bond length and bond angles are taken to be those which minimize the ground state energy, and the binding energies are calculated by taking the difference between the ground state energy calculated at the disassociation limit ($>$ 10 Å) with the minimum energy. We have also used each method to produce the potential energy curves of each molecule. To be consistent, we use the same number of qubits before quadratization for both methods. We compare the results we obtain to those of the restricted Hartree-Fock (RHF) and complete active space configuration interaction (CASCI) methods \cite{RHF+CASCI}. Note that the CASCI method is exact for the chosen basis set and active space, and it is equivalent to the full configuration interaction (FCI) method when the active space includes all spin-orbitals. We therefore use the CASCI method to measure the accuracy of the XBK and QCC methods.

The scalability of the XBK and QCC methods depends on the number of post-quadratization qubits needed to run the methods and the total computation time. The largest quantum annealer currently available is D-Wave's new Advantage system, which has over 5000 qubits with 15-qubit connectivity. To compare the computational costs of each method, we plot how the number of post-quadratization qubits scales with the number of pre-quadratization qubits and thus with the parameters of each method. We also plot the time required to compute a single ground state energy versus the number of pre-quadratization qubits to gauge how the computation time scales with the size and accuracy of the calculations.

For all calculations, we utilize the PySCF and OpenFermion modules to construct the relevant operators and to calculate the RHF and CASCI energies \cite{PySCF, OpenFermion}. The total number operator is used to fix the number of electrons as in Eq. \ref{PenaltyTerm}, and the Bravyi-Kitaev transformation is used to map the fermionic Hamiltonian to Pauli operators. We use D-Wave's Ocean Software to quadratize the Hamiltonians and embed them on the annealer \cite{D-Wave, Choi2008}. D-Wave's Advantage quantum annealer is used for the time-sensitive calculations; however, due to the limited computational time available on the D-Wave we use the simulated annealer available through the Ocean Software for the remaining calculations. All classical computations are done on an AMD Ryzen 7 1700X Eight-Core processor running at 3400 MHz. We use the L-BFGS-B algorithm to perform the continuous optimization in the QCC method \cite{Byrd1995}. The code we used for this project can be found at \url{https://github.com/jcopenh/Quantum-Chemistry-with-Annealers}.

\subsection{\label{H3+}Trihydrogen Cation}

As the most common ion in the universe, H$_3^+$ provides an interesting subject to test the efficacy of the XBK and QCC methods for ions. The nuclei of H$_3^+$ form an equilateral triangle with an H-H equilibrium bond length of about 0.9 Å. As far as we are aware, this is the first time H$_3^+$ will be modeled on a quantum annealer. For H$_3^+$, we use the STO-6G basis set with all 6 spin-orbitals, and the Hamiltonian is written using 4 qubits after applying symmetry reductions. We were able to run the XBK method with up to $r=4$, necessitating 16 pre-quadratization qubits. For the QCC method, we set $N_{ent}=4$ and folded the $\theta_i$ and $\varphi_i$ variables once and the $\tau_k$ variables twice, again needing 16 qubits.

\begin{figure}[t]
    \includegraphics[width=.48\textwidth]{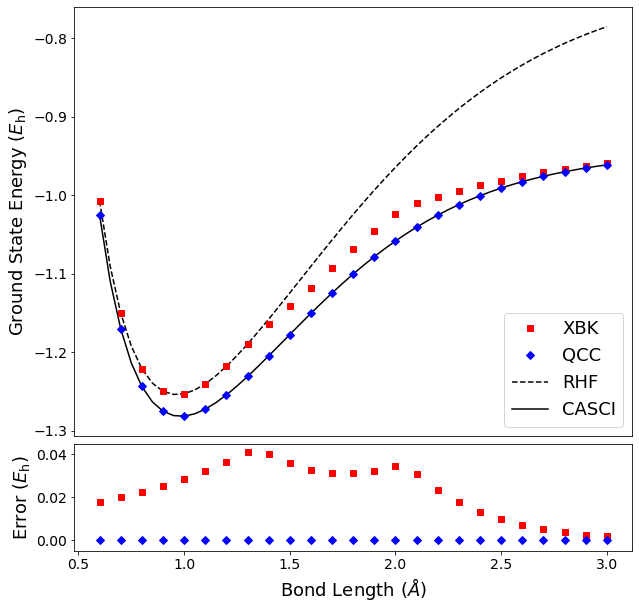}
    \caption{Potential energy curve for H$_3^+$ created by symmetrically varying the H-H bond lengths. The difference between the XBK and QCC energies and those calculated using the CASCI method are plotted below.}
    \label{H3+_PES}
\end{figure}

\begin{table}[b]
    \centering
    \begin{tabular}{ |wc{1.5cm}|wc{1.5cm}|wc{1.5cm}| } 
        \hline
        Method & BE ($E_\mathrm{h}$) & BL (Å) \\
        \hline
        XBK    & 0.312 & 0.965 \\
        QCC    & 0.339 & 0.984 \\
        RHF    & 0.560 & 0.965 \\
        CASCI  & 0.339 & 0.984 \\
        \hline
    \end{tabular}
    \caption{Binding energy and bond length of H$_3^+$ calculated using various methods.}
    \label{H3+ Data}
\end{table}

The potential energy curve associated with symmetrically stretching the H-H  bonds is shown in Fig. \ref{H3+_PES}. Here, the CASCI energies are exact for the STO-6G basis since all spin-orbitals are included in the active space. At $r=4$, the XBK method is able to find energies lower than the RHF energies except near the equilibrium length, but is outperformed by the QCC method, which consistently finds the ground state energies to within chemical accuracy ($<$ 0.002 Hartree). Table \ref{H3+ Data} shows the values for the binding energy and bond length of H$_3^+$ calculated using the various methods. The XBK method shows improvement over the RHF method, while the QCC method gives the exact values.

\begin{figure}[t]
    \includegraphics[width=.48\textwidth]{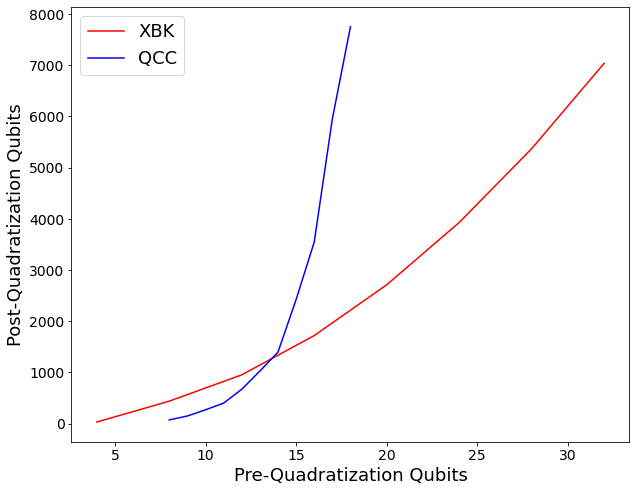}
    \caption{Number of post-qudratization qubits required to run each method versus the number of pre-quadratization qubits for H$_3^+$ with a bond length of 1.2 Å.}
    \label{H3+_qubits}
\end{figure}

\begin{figure}[t]
    \includegraphics[width=.48\textwidth]{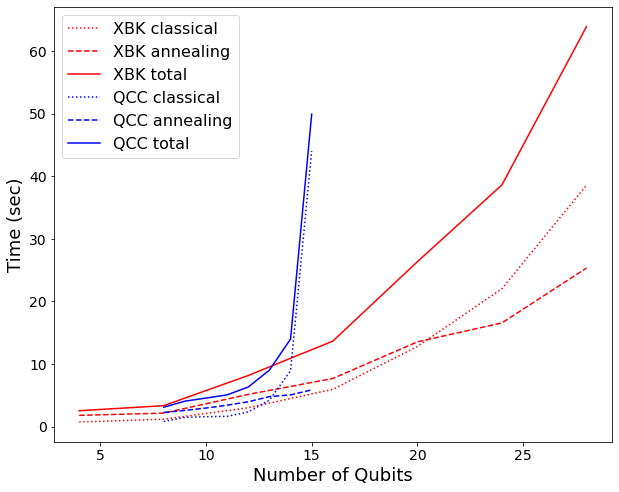}
    \caption{Breakdown of the computation times of the XBK and QCC methods versus the number of pre-quadratization qubits for H$_3^+$ with a bond length of 1.2 Å.}
    \label{H3+_time}
\end{figure}

In Fig. \ref{H3+_qubits} we plot the qubit scaling for the H$_3^+$ molecule at a bond length of 1.2 Å. For the XBK method the value of $r$ is varied, while for the QCC method we vary $N_{ent}$ and fold all three variables once. Note that for H$_3^+$ with the chosen settings the number of pre-quadratization qubits will go as $4r$ for the XBK method and $8+N_{ent}$ for the QCC method. Fig. \ref{H3+_time} shows the time scaling of the H$_3^+$ molecule at 1.2 Å. The computation time averaged over 5 runs is broken into two components. The "classical" time includes the time required to convert the electronic Hamiltonian to Pauli operators, find an embedding for the Ising model, and any other procedure performed on the classical processor. The "annealing" time is the total amount of time the quantum annealer spends on the calculation, which includes the physical annealing as well as several other steps, as reported by D-Wave's qpu$\_$access$\_$time variable. For reference, the convergence time of the CASCI method was 0.23 sec. 

As Figs. \ref{H3+_qubits} and \ref{H3+_time} demonstrate, the resource requirements of the XBK and QCC methods do not scale well with the parameters of each method. The number of post-quadratization qubits required to simulate H$_3^+$ surpasses what is available on D-Wave's Advantage system after $r=6$ for the XBK method and $N_{ent}=8$ for the QCC method. Similarly, the computation time of each method increases exponentially with the number of qubits. Since more qubits are required to accurately calculate the energies of larger molecular systems, this result indicates that neither method will scale well with system size.

\subsection{\label{O_2}Water Molecule}

The water molecule is a slightly larger molecule than H$_3^+$ and thus provides a better test for the limits of the XBK and QCC methods. We use the 6-31G basis set to construct the fermionic Hamiltonian, but restrict the active space to just 8 spin-orbitals and 4 electrons due to computational constraints. Since the active space does not include all of the spin-orbitals, the CASCI method is not exact. After applying symmetry reductions, the Hamiltonian can be written using 5 qubits. The XBK method was only run with $r=2$, while for the QCC method $N_{ent}=5$ and the $\theta_i$ and $\tau_k$ variables were folded once; thus, both methods used 10 pre-quadratization qubits.

\begin{table}[b]
    \centering
    \begin{tabular}{ |wc{1.5cm}|wc{1.5cm}|wc{1.5cm}|wc{1.5cm}| } 
        \hline
        Method & BE ($E_\mathrm{h}$) & BL (Å) & BA (\degree) \\
        \hline
        XBK    & 0.257 & 0.954 & 111.2 \\
        QCC    & 0.262 & 0.960 & 110.5 \\
        RHF    & 0.602 & 0.954 & 111.2 \\
        CASCI  & 0.265 & 0.968 & 109.4 \\
        \hline
    \end{tabular}
    \caption{Binding energy, bond length, and bond angle of H$_2$O calculated using various methods.}
    \label{H2O Data}
\end{table}

\begin{figure}[t]
    \includegraphics[width=.48\textwidth]{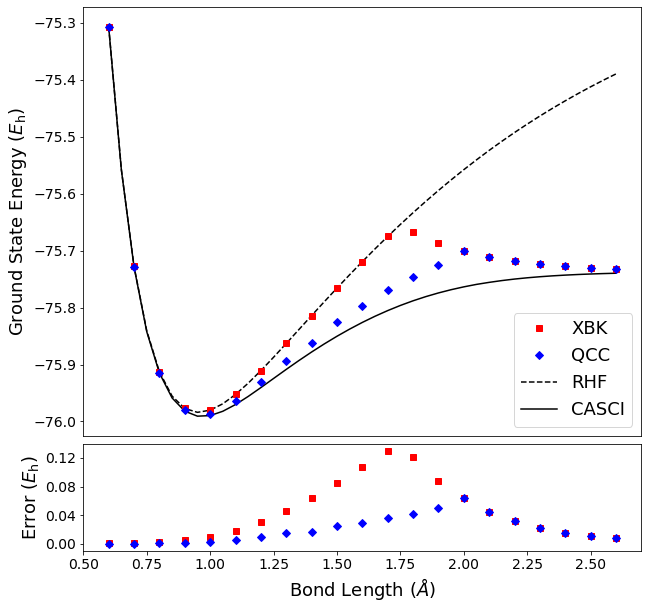}
    \caption{Potential energy curve for H$_2$O created by symmetrically varying the O-H bond lengths with fixed bond angle of 104.48\degree. The difference between the XBK and QCC energies and those calculated using the CASCI method are plotted below.}
    \label{H2O_PES}
\end{figure}

\begin{figure}[t]
    \includegraphics[width=.48\textwidth]{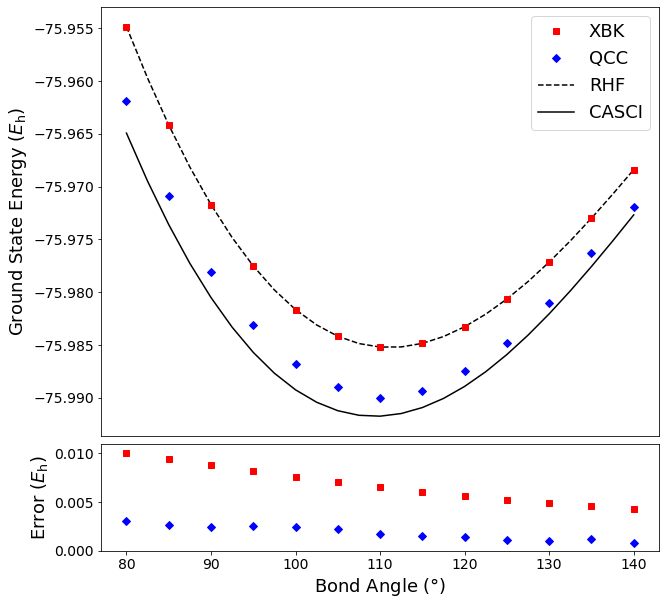}
    \caption{Potential energy curve for H$_2$O created by varying the angle between the O-H bonds with fixed bond lengths of 0.9578 Å. The difference between the XBK and QCC energies and those calculated using the CASCI method are plotted below.}
    \label{H2O_BA_PES}
\end{figure}

Fig. \ref{H2O_PES} shows the potential energy curve created by symmetrically stretching the O-H bonds, keeping the bond angle at a constant 104.48\degree. At $r=2$, the XBK method follows the RHF curve near the equilibrium point, but then quickly converges to the CASCI curve. The QCC method, meanwhile, consistently finds energies below the RHF curve, with the most accurate results again found near the equilibrium point and in the asymptotic region. In Fig. \ref{H2O_BA_PES} we show the potential energy curve created by varying the bond angle with the bond length set to 0.9578 Å. In the region analyzed with $r=2$, the XBK method is unable to find energies lower than the RHF method while the QCC method demonstrates a marked improvement in accuracy at every bond angle.

The calculated values of the binding energy, bond length, and bond angle of H$_2$O are shown in Table \ref{H2O Data}. Since the XBK method returned the RHF energies near the equilibrium point, the calculated bond length and angle are the same as in the RHF method. The binding energy is closer to that given by the CASCI method due to the asymptotic behavior of the XBK method. The QCC method nears chemical accuracy for the binding energy, and shows improvement for the bond length and angle.

\begin{figure}[t]
    \includegraphics[width=.48\textwidth]{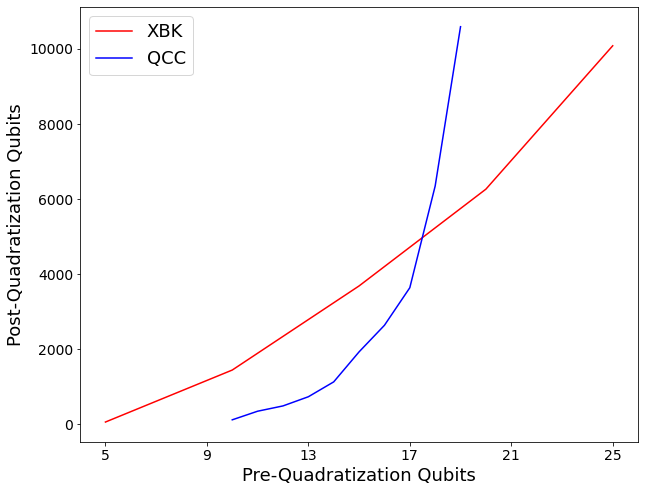}
    \caption{Number of post-qudratization qubits required to run each method versus the number of pre-quadratization qubits for H$_2$O with a bond length of 1.6 Å.}
    \label{H2O_qubits}
\end{figure}

\begin{figure}[t]
    \includegraphics[width=.48\textwidth]{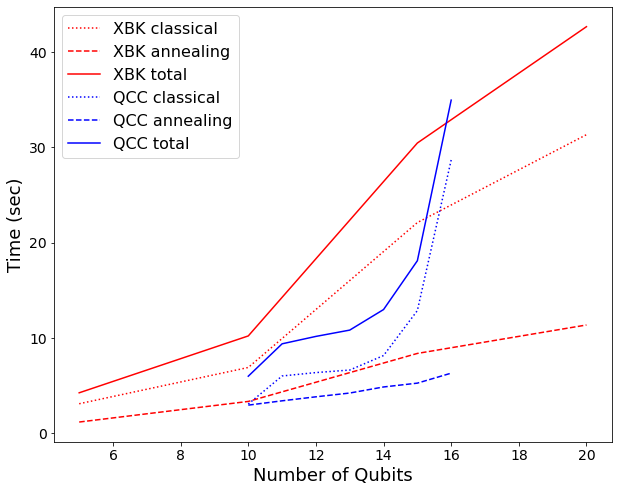}
    \caption{Breakdown of the computation times of the XBK and QCC methods versus the number of pre-quadratization qubits for H$_2$O with a bond length of 1.6 Å.}
    \label{H2O_time}
\end{figure}

Figs. \ref{H2O_qubits} and \ref{H2O_time} show the qubit and time scaling, respectively, of H$_2$O with a bond length of 1.6 Å. For H$_2$O the number of pre-quadratization qubits is $5r$ for the XBK method and $10+N_{ent}$ for the QCC method. Fig. \ref{H2O_qubits} thus indicates that the number of post-quadratization qubits required to simulate H$_2$O surpasses what is available on the Advantage system after $r=3$ and $N_{ent}=7$. The scaling of both metrics is similar to the results for H$_3^+$, indicating that as the number of pre-quadratization qubits increases the resource requirements increase rapidly. The CASCI method required just 0.38 sec.

\section{\label{Conclusion}Concluding Remarks}

As demonstrated in the previous sections, it is possible to perform electronic structure calculations by using quantum annealers in tandem with a classical computer. However, we see that the time required to run the XBK and QCC methods is much greater than their classical counterparts. The reasons for this are twofold: first, the requirement that the problem Hamiltonian mapped on the annealer be 2-local results in an exponential increase in the number of qubits as ancillary qubits are introduced during quadratization, necessitating more qubits on the annealer and leading to longer run-times. Second, each method requires extensive, time-consuming use of the classical computer, erasing any potential quantum speedup.

The XBK method requires a large number of pre-quadratization qubits to achieve results much better than the RHF method. Since the number of post-quadratization qubits increases rapidly with the number of pre-quadratization qubits, the XBK method thus quickly surpasses the number of qubits available on modern annealers, making the accurate simulation of larger molecules difficult. The QCC method demonstrates improvement over the XBK method by achieving a greater level accuracy using fewer qubits. Unfortunately, the QCC method leans on the classical computer more heavily by using it to perform the bulk of the optimization. Hence, the computation time of the QCC method scales largely the same as the classical optimization algorithm used. In addition, the number of post-quadratization qubits increases even faster for the QCC method, such that using the method for systems much larger than H$_2$O is only possible with very few entanglers and minimal folding, resulting in very little improvement over the RHF energies. For these reasons, neither method is able to accurately simulate all but the smallest of molecules.

There are a few potential avenues through which to improve the speed and accuracy of quantum chemical simulations on quantum annealers. From the software side, new methods could be developed to map the problem Hamiltonian to the Ising model using fewer qubits. This could involve either a more efficient transformation to z-type Pauli operators or better techniques for utilizing molecular symmetries. A potential hardware solution, which would likely be much more beneficial, would be to develop large-scale annealers implementing non-stoquastic Hamiltonians \cite{Biamonte2008, Barends2016, Vinci2017, Grant2020, Ozfidan2020}. Such an annealer would be universal and could utilize methods such as Hamiltonian gadgets to quadratize the Hamiltonian \cite{Babbush2013, Cao2015}. This would enable one to avoid the introduction of large numbers of ancillarly qubits during the quadratization process and to more efficiently simulate larger molecules.

\begin{acknowledgments}
We would like to thank Raunaq Kumaran for thoughtful discussions. We would like to thankfully acknowledge financial support from the Discovery Park Undergraduate Research Internship Program (DURI) at Purdue and from the NSF under the grants PHY-1255409 and CHE-1900301.
\end{acknowledgments}

\section*{Data Availability}
The data that support the findings of this study are available from the corresponding author upon reasonable request.

\nocite{*}
\bibliography{references}
\end{document}